\documentclass[%
reprint,
superscriptaddress,
 amsmath,amssymb,
prd,
]{revtex4-1}

\usepackage{graphicx}
\usepackage{dcolumn}
\usepackage{bm}
\usepackage{hyperref}
\usepackage{epsfig,amsmath,amssymb,verbatim,mathrsfs,array,layout,textcomp,amssymb,latexsym,slashed,graphicx,booktabs,color,mathtools,tikz}
\usepackage{siunitx}
\usepackage{multirow}

\def\beq{\begin{equation}}
\def\eeq{\end{equation}}

\begin{document}

\title{Diamond Detectors for Direct Detection of Sub-GeV Dark Matter}

\author{Noah Kurinsky}\affiliation{Fermi National Accelerator Laboratory, Batavia, IL 60510, USA} \affiliation{Kavli Institute for Cosmological Physics, University of Chicago, Chicago, IL 60637, USA}
\author{To Chin Yu}\affiliation{Department of Physics, Stanford University, Stanford, CA 94305, USA}
\affiliation{SLAC National Accelerator Laboratory, 2575 Sand Hill Road, Menlo Park, CA 94025, USA}
\author{Yonit Hochberg}\affiliation{Racah Institute of Physics, Hebrew University of Jerusalem, Jerusalem 91904, Israel}
\author{Blas Cabrera}\affiliation{Department of Physics, Stanford University, Stanford, CA 94305, USA}


\begin{abstract}
We propose to use high-purity lab-grown diamond for the detection of sub-GeV dark matter. Diamond targets can be sensitive to both nuclear and electron recoils from dark matter scattering in the MeV and above mass range, as well as to absorption processes of dark matter with masses between sub-eV to 10's of eV.
Compared to other proposed semiconducting targets such as germanium and silicon, diamond detectors can probe lower dark matter masses via nuclear recoils due to the lightness of the carbon nucleus. The expected reach for electron recoils is comparable to that of germanium and silicon, with the advantage that dark counts are expected to be under better control. Via absorption processes, unconstrained QCD axion parameter space can be successfully probed in diamond for masses of order 10~eV, further demonstrating the power of our approach. 

\end{abstract}

\maketitle


\section{Introduction}\label{sec:intro}

The identity of the dark matter (DM) in the universe is one of the most pressing puzzles of modern day physics. Guided by the Weakly Interacting Massive Particle (WIMP) paradigm,  experimental efforts have focused for decades on tracking down DM at the GeV mass scale and above. As sensitivity to this mass range continues to increase with large scale detectors and reduced thresholds, the lack of observation of WIMP particles stresses the importance and timeliness of searching for lighter DM beyond the WIMP.

Indeed, recent years have seen a surge in ideas for sub-GeV DM detection, including the use of atomic ionization~\cite{Essig:2011nj}, semiconductors such as germanium (Ge) and silicon (Si)~\cite{Essig:2011nj,Essig:2012yx,Graham:2012su}, scintillators, color centers~\cite{Budnik:2017sbu}, two-dimensional targets such as graphene~\cite{Hochberg:2016ntt} and carbon nanotubes~\cite{Cavoto:2017otc}, superconductors~\cite{Hochberg:2015pha,Hochberg:2015fth,Hochberg:2016ajh}, Dirac materials~\cite{Hochberg:2017wce}, polar crystals~\cite{Knapen:2017ekk,Griffin:2018bjn} and superfluid helium~\cite{Schutz:2016tid,Knapen:2016cue,hertel}. Some of these proposals make use of DM-electron interactions, some of DM-nucleon interactions, while some are sensitive to both.

Here we propose the use of diamond detectors for sub-GeV DM. Such detectors can probe both electron and nuclear recoils from DM scattering in the MeV and above mass range, as well as sub-eV to 10's of eV DM masses via absorption processes. 

Compared to other proposals, diamond detectors have several advantages. First, the light nucleus of carbon enables detection of lower DM masses compared to other targets for nuclear recoils. Second, compared to other semiconductor targets, diamond has excellent isotopic purity; more energetic and long-lived phonon modes with higher velocities; long phonon mean free paths, which should allow for larger crystals; and is radiation hard. Diamond targets can also hold large electric fields ($> 20$~MV/cm). Finally, the absence of electrical impurity states below $0.5$~eV is suggestive of low dark count rates. In addition, 
much of the current R\&D and technology developed for silicon and germanium targets can be ported over to diamond with minimal modification, placing diamond detectors in an excellent position to broadly probe and detect DM in the near future. 

This paper is organized as follows. In Section~\ref{sec:med} we discuss aspects of diamond as a detector medium, including particle interactions in diamond, charge and phonon collection efficiency, and resolution. In Section~\ref{sec:backgrounds} we briefly discuss potential limiting backgrounds, as compared to Si and Ge. Our projected reach for dark matter absorption, electron recoils and nuclear recoils is presented in Section~\ref{sec:results}. We conclude with a discussion in Section~\ref{sec:disc}.

\section{Diamond as a Detection Medium}\label{sec:med}
\begin{figure}[ht]
    \centering
    \includegraphics[width=3.25in]{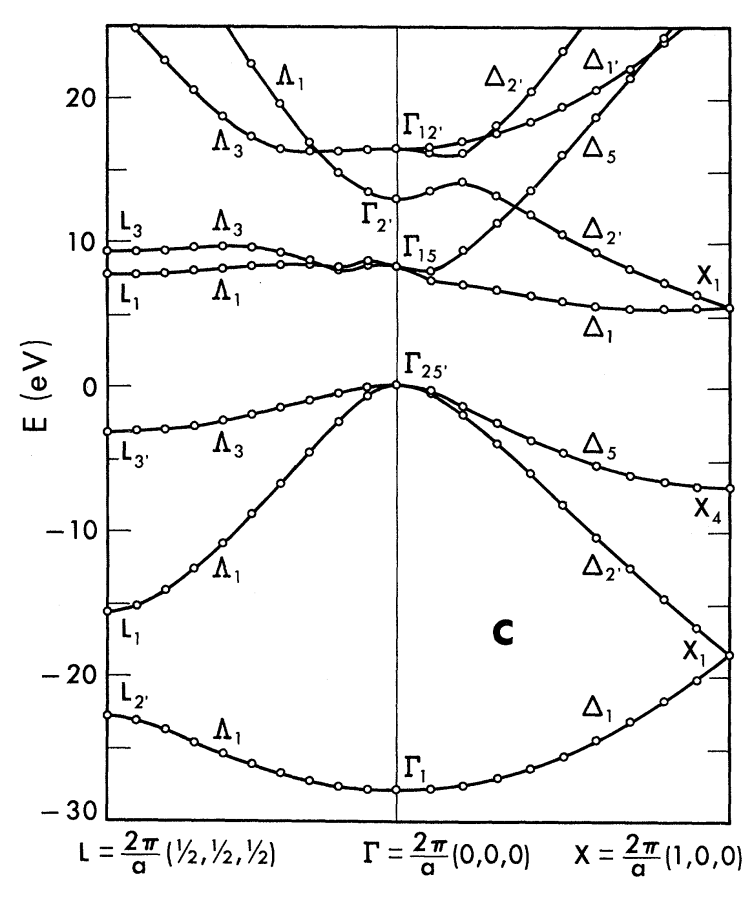}
    \caption{Band structure of diamond in the reduced-zone scheme, reproduced from Ref.~\cite{saslow}. }
    \label{fig:bs}
\end{figure}

Diamond, like silicon and germanium, is a semiconductor with a tetrahedral lattice symmetry and an indirect bandgap~\cite{saslow,dolling,VavilovDiamond,Nava}. The strong nature of the carbon-carbon covalent bonds leads to a larger bandgap energy (5.4 eV) and more energetic optical phonon modes ($\sim$160-190~meV) than Si, but the p-orbital dominated band structure is qualitatively very similar to Si, in that the energy valleys are highly anisotropic, and the minima occur along the X-valleys in the Brillouin zone (see Fig.~\ref{fig:bs}). 

These properties theoretically give high-purity diamond very high carrier lifetime, thermal conductivity, and high resistivity to much higher temperatures than either Si or Ge. The large bandgap also makes diamond transparent to IR and optical photons, and the dominant impurity states all have binding energies much larger than that found in Si or Ge, giving diamond much lower IR-induced dark counts. Some lattice properties of diamond, as compared to Si and Ge, can be found in Table~\ref{tab:properties}.

\begin{table*}[t]
\begin{tabular}{| c | l | c | c | c |}
\hline
Parameter & \multicolumn{1}{|c|}{Description} & Diamond (C) & Si & Ge \\ 
\hline
Z & Atomic number & 6 & 14 & 32 \\
A & Average atomic mass & 12.01 & 28.09 & 72.64 \\
  & Stable isotopes & 12,13 & 28,29,30 & 70,72,73,74 \\
  & Natural radioactive isotopes & 14 & 32 & 76 \\
\hline
a (A) & Lattice spacing & 3.567 & 5.431 & 5.658 \\
N (cm $^{-3}$) & Number density & $1.76\times10^{23}$ & $5\times10^{22}$ & $4.42\times10^{22}$\\
$E_{\rm gap}$ (eV)& Bandgap Energy & 5.47 & 1.12 & 0.54\\
$E_{eh}$ (eV) & Average energy per $\mathrm{e^{-}h^{+}}$ pair & $\sim$13~\cite{Ziaja} & 3.6-3.8~\cite{Ziaja,CDMSSensitivity} & 3.0~\cite{CDMSSensitivity} \\
$\epsilon_r$ & Relative permittivity & 5.7 & 11.7 & 16.0 \\
$\Theta_{\rm Debye}$ (K) & Debye temperature & 2220 & 645 & 374 \\
$\hbar\omega_{\rm Debye}$ (meV) & Debye energy & 190 & 56 & 32 \\
$\hbar\omega_{\rm TO}$ (meV) & Transverse optical phonon energy & 141 & 59 & - \\
$\hbar\omega_{\rm LO}$ (meV) & Longitudinal optical phonon energy& 163 & 63 & 37 \\
$c_s$ (m/s) & Average phonon speed & 13360 & 5880 & 3550 \\
$v_{d,sat}$, $\mathrm{e^{-}}$ (m/s) & Electron saturation velocity & $\sim2\times 10^{5}$ & $1.35\times 10^{5}$ & $1.2\times 10^{5}$ \\
$E_{\textrm{Bd}}$ (MV/cm) & Dielectric breakdown field & $>$20~\cite{Landstrass} & 0.3 & 0.1  \\ 
\hline
        $\ell$ (cm) & Phonon mean free path & 25.6  & 11.4 & 6.9 \\
        $\tau_{\mathrm{life}}$ ($\mathrm{\mu s}$) & Phonon Lifetime & 19.2  & 19.4 & 19.5 \\
        $f_{\mathrm{W}}$ & Phonon Transmission (W) &  80.6\% & 42.1\%  & 62.6\%  \\
        $c_sf_{\mathrm{W}}$ (m/s) & Effective speed (W) &  10773 & 2473 & 2221 \\
        $f_{\mathrm{Al}}$ & Phonon Transmission (Al) & 63.7\% & 90.0\% & 80.9\% \\
        $c_sf_{\mathrm{Al}}$ (m/s) & Effective speed (Al) & 8513 & 5290 & 2872 \\
\hline
\end{tabular}
\caption{{\it First and second sections:} material properties of diamond, Si, and Ge (from Refs.~\cite{Jacoboni,Neves2001,Isberg} unless otherwise stated). {\it Last section:} Phonon figures of merit in the three materials at low temperatures ($\lesssim30$mK) for a mm-scale crystal. See text for more details.}
\label{tab:properties}
\end{table*}

The semiconducting nature of diamond was well established theoretically around the time that Si and Ge were first being used to create transistors~\cite{saslow}, but the development of diamond electronics was slowed by the cost of scaling diamond as a technology, as well as the lack of an adequate donor impurity, which rendered all early diamond devices implicitly n-type~\cite{VavilovDiamond}. Despite these limitations, natural diamond was successfully used in the late 1970s to produce an avalanche-type particle detector~\cite{Kozlov}\footnote{In this case, type IIa diamond, which is the purest grade of natural diamond, was used. The limitation of this experiment was finding a natural diamond with sufficient purity, given the variability between diamonds. Type IIa diamonds have nitrogen as their primary impurity.}. As sufficiently high-purity synthetic crystals have become available in the past two decades~\cite{Isberg}, diamond has shown success both as an X-ray detector and as a particle detector for high-intensity nuclear radiation environments (see Ref.~\cite{Tarun} and references therein). In this time, the field of diamond electronics has become much more mature~\cite{Isberg,DiamondElectronics}, and as industry has begun to produce lower-cost synthetic diamond substrates, these devices have come into broader use in particle physics, most recently as upgrades to both the ATLAS and CMS detectors~\cite{HEPdiamond}. Lattice vacancies in diamond have also become interesting to the quantum sensing community as potential qubit storage media, which may allow the use of diamond as a directional dark matter detector when used in conjunction with conventional technologies~\cite{Rajendran}.

Here we explore the possibility of producing diamond detectors with sub-eV energy thresholds by employing cryogenic phonon and charge readouts similar to those used by the SuperCDMS, CRESST and EDELWEISS experiments to achieve eV-scale thresholds in Si~\cite{Romani}, Ge~\cite{kurinsky,EdelweissRnD}, $\mathrm{CaWO_3}$~\cite{CRESSTIII}, and $\mathrm{Al_2O_3}$ (sapphire)~\cite{strauss}, respectively. We will touch on the ionization yield model for diamond, and discuss relevant aspects of the charge-phonon dynamics, before discussing reference designs capable of achieving the performance shown in our sensitivity projections later in this paper. 
The theorist interested primarily in the expected reach of diamond into dark matter parameter space can move directly to Section~\ref{sec:results}.

\subsection{Particle Interactions in Diamond}

In the energy range of interest for sub-GeV dark matter searches ($\sim <$1~keV), there is a large difference in detector response between nuclear and electronic recoils. For a pure calorimeter, the only quantity of interest is the recoil energy, which does not depend on the partition of energy into phonons (often called displacement energy) and free charge carriers (ionization energy). For many dark matter detectors, however, the difference in this partition is a useful way to discriminate between types of detector interactions, and the amount of discrimination ability in this channel is an important property of a potential diamond detector.

At very high energies, on the MeV-scale, the vast majority of the initial energy lost in the interaction between the incident particle and the diamond substrate is lost to charge production~\cite{Ziaja,LindhardDiamond}. Ref.~\cite{CanaliRadDet} demonstrated that $\alpha$ particles with energies in the MeV range and $\beta$ particles down to energies of $\sim$~200~keV produce an average of one electron-hole pair per 13~eV of recoil energy in diamond. In this paper we refer to this energy as $E_{eh}$. Subsequent follow up experiments with proton, neutron, ion and electron beams have shown this is true for a wide range of particles and energies (see {\it e.g.} Refs.~\cite{LindhardDiamond,Pernegger,Kagan,Tarun}). Calculations of the Lindhard partition function $y(E_R)$ in diamond, defined as
\begin{equation}
y(E_R)=\frac{E_{\rm ionization}}{E_{\rm recoil}}=\frac{E_{eh}n_{eh}}{E_{\rm recoil}},
\end{equation}
where $n_{eh}$ is the number of electron-hole pairs produced, suggest that the energy partition is very similar to that found in Si~\cite{LindhardFactors,LindhardDiamond}. From these studies, we see that for all nuclear recoils with energies above $\sim$1~MeV, we can expect both electronic and nuclear recoils to produce the same ratio of charge carriers to phonon energy, $y(E_R)=1$. The ratio of electron-hole pair production energy $E_{eh}$ to bandgap energy $E_{\rm gap}$ means that a charge yield of 1 is slightly misleading, and that the actual energy stored in the charge system is given by the ratio $E_{\rm gap}/E_{eh}$, the remainder of the energy shed by the initial charge carriers into the phonon system.

The fraction of energy dissipated in nuclear recoils by charge production drops to around 50\% of its high-energy value by 10~keV, and is expected to exactly match that of Si by $\sim$100~eV~\cite{LindhardDiamond}. This comes with the caveat that the ionization yield for nuclear recoils in both Si and diamond are essentially unmeasured below 1~keV, which is an outstanding challenge for the upcoming generation of dark matter searches~\cite{CDMSSensitivity}. 

The electron-recoil response of diamond, however, is well validated over our entire energy range of interest, maintaining a value of $y(E_R)=1$ across this range. There is a large community interested in the potential application of diamond to UV and X-ray photon detection~\cite{UVDet01}, and recent work has demonstrated full charge collection (consistent with the 13~eV per pair energy) for a range of photons both near the bandgap~\cite{UVDet12} and in soft X-rays, for energies between 200~eV and 2~keV~\cite{Keister}. These measurements are well matched by models of secondary electron cascades in Si~\cite{Ziaja}. The lack of explicit calibration in the energy range between $\sim$10~eV and $\sim$200~eV can be attributed to the difficulty in producing calibration sources in this energy range, and in subsequently propagating these photons to the bulk of the detector through electrodes (which is also a challenge for similar Si and Ge detectors). In the case of a dark matter search, this intrinsic shielding is actually a highly desired property, though it will continue to present a challenge for calibration. Compton recoils from higher energy gamma rays show promise as a reliable means to calibrate the energy scale and electron-recoil ionization yield for future low-threshold detectors, including diamond.

We note that interactions in high-purity diamond are not limited to energies above the bandgap. Nuclear recoils can excite phonons of arbitrarily low energy, however such small momentum transfers will not generate a crystal defect since the displacement energy of a nucleus from its lattice site is ~40 eV~\cite{LindhardDiamond}. But sub-gap interactions can kinematically excite a nucleus, producing athermal phonons until the system returns to thermal equilibrium. Photon scattering can also occur through multi-phonon excitations for phonon energies up to the Debye temperature of the crystal~\cite{Hochberg:2016sqx}, and through coherent scattering with an atom~\cite{robinson}. In addition, impurities in the lattice can create sub-gap bound states, which we will consider as sources of backgrounds in Section~\ref{sec:backgrounds}.

Given an estimate for the charge and phonon production from an interaction of known energy, it is important to consider the long-term stability of these excitations, which will determine how useful they will be for particle detection. In the next two subsections, we consider the long-term stability of charge carriers and phonons in high-purity diamond, and consider the efficiency and precision with which these excitations can be collected and measured based on recent developments in low-threshold detector technologies.

\subsection{Charge Readout}
\subsubsection{Charge Collection Efficiency}

The first demonstration of sufficient charge mobility in synthetic diamond, which touched off the recent progress in development of diamond-based sensors, showed a charge carrier lifetime in excess of 2~$\mu {\rm sec}$~\cite{Isberg}, which allows us to get a sense for the charge collection efficiency (CCE) possible in a diamond device of different sizes. For a field strength of 1-10 kV/cm, charge drift velocity is on the order of $10^7$ cm/sec or 10 cm/$\mu {\rm sec}$~\cite{Nava}. Thus a charge carrier has a mean free path, at room temperature, of at least 10~cm; this is in fact the maximum collection distance measured by Ref.~\cite{Isberg} for high-quality chemical vapor deposition (CVD) diamonds at room temperature, though it is by no means a fundamental limit. It is likely that as CVD diamond continues to improve in purity, these collection lengths will only increase.

If we note that carrier lifetime and drift velocity both increase exponentially at lower temperatures~\cite{Nava}, we infer that high-quality CVD diamond can in principle offer perfect charge collection over macroscopic crystal sizes; certainly larger than any commercially available CVD crystals. For example, Ref.~\cite{Tarun} demonstrated that charge collection in diamond is primarily determined by nitrogen defects by observing CCE as a function of time in a high radiation environment. Thus sufficiently pure, low-temperature CVD diamond should provide highly efficient charge collection, which can be enhanced, and the bandwidth increased, by cooling the diamond to moderate cryogenic temperatures ($\sim 4$~K or higher, possibly up to liquid nitrogen temperatures).

\subsubsection{Charge Resolution}

The simplest extension of past work producing diamond detectors (see {\it e.g.} Refs.~\cite{Kozlov,CanaliRadDet,Neves2001,Tarun,Keister}) is to consider how advances in Si and Ge charge detectors can be leveraged to produce a charge-quantum sensitive diamond detector. 

We first consider a simple model of the solid-state ionization ``chamber". In this model, a single non-doped monolithic crystal is sandwiched between two electrodes, and a bias is applied to drift the electron-holes pairs to the electrodes when an interaction occurs in the crystal. It has already been demonstrated that full charge collection (without losses due to trapping or recombination), even without explicit cooling, can be achieved for events occurring in the bulk of a high-purity CVD crystal~\cite{Keister}. 

The only challenge then is to couple a sufficiently pure substrate to a readout with adequate resolution. In addition, the characteristic collection time is given by $\tau_q \sim \eta/v_d(E,T)$, where $\eta$ is the crystal thickness and $v_d(E,T)$ is the drift velocity at a given electric field strength and temperature. CVD diamonds, with thickness on the order of $0.1$~cm and drift velocities of $\sim 10^7$~cm/sec can expect to have $\tau_q \lesssim 10$~nsec~\cite{Nava}. The charge collection will therefore almost always be faster than the charge readout circuit, and for small crystals we can ignore charge collection time to a very good approximation.

For the ionization chamber model, there is negligible current noise contributed by the system itself, given that it is completely frozen out. Any thermionic emission of charge shows up as a signal event rather than contributing to the charge noise budget, as shown in Ref.~\cite{Romani}. The standard way of detecting charge is using an integrator circuit. The minimum resolution of a charge integrating readout is completely determined by the noise properties of the amplifier, the bias circuit, and the capacitance of the detector ($C_{\rm det}$) and amplifier ($C_{\rm in}$) (see {\it e.g.} Ref.~\cite{shuttThesis}):
\begin{equation}
\sigma_{q} \ge \frac{N_{v}(C_{\rm det}+C_{\rm in})}{\epsilon_q \sqrt{\tau}},
\end{equation}
where $N_{v}$ is assumed to be a flat voltage noise spectral density of the amplifier in $V/\sqrt{\rm Hz}$, $\epsilon_q$ is the CCE and $\tau$ is the response time of the detector and readout. For an integrator, the readout time $\tau$ is determined by the rate at which the input is drained by some bias resistor $R_b$, and thus $\tau = R_b (C_{\rm det}+C_{\rm in})$. 

In principle, the readout time can be arbitrarily long by taking $R_b\rightarrow \infty$, but in practice the readout time is always limited by either pileup events or by rising noise at low frequency (often called 1/f noise), presumably caused by stochastic transitions in two-level systems. Recent work using High Electron-Mobility Transistors (HEMTs) by the SuperCDMS and EDELWEISS collaborations demonstrated a charge readout with a noise corner of around 4~kHz ($\tau\approx 40\;\mathrm{\mu}$sec) and a white noise floor of $N_{v}\sim$~0.2~nV/$\sqrt{\mathrm{Hz}}$, resulting in a charge resolution of $\sim35$ electron-hole pairs for $C_{\rm det}\approx 150$~pF and $C_{\rm in}\approx 100$~pF~\cite{Phipps}. These amplifiers can be produced with variable $C_{\rm in}$ down to $<$10~pF, and it has been shown that the 1/f noise scales as $C_{\rm in}^{-1/2}$ while the white noise is invariant to gate capacitance~\cite{dong}. This means that $\tau \propto C_{\rm in}^{1/2}$, and the charge resolution scales as
\begin{equation}
\sigma_{q}\approx (35\;\mathrm{e^{-}h^{+}\;pairs})\frac{(C_{\rm det}+C_{\rm in})/(250~\mathrm{pF})}{(C_{\rm in}/100~\mathrm{pF})^{1/4}}.
\end{equation}
A typical CVD diamond crystal with a 4~mm$\times$4~mm face and 0.5~mm thickness has a capacitance of 2-5~pF; the reduction in capacitance comes both from the smaller physical size, and from the lower relative permittivity of diamond as compared to Si and Ge (see Table~\ref{tab:properties}). Matching the input gate capacitance to the detector capacitance therefore predicts a charge resolution of 1--3 electron-hole pairs (depending on the actual crystal capacitance) for ideal operating conditions in an ionization chamber mode. This is likely near the best achievable charge resolution for a cryogenic diamond detector of the proposed size using current readout electronics, and is nearing sub-electron resolution.

In contrast to the single-channel ionization chamber model of charge readout, we can consider the more common paradigm of massively parallel readout used in CCDs. Recent advances in Si CCD technology have enabled single-read resolution per pixel of 2~$e^{-}$, and allow for multiple reads on a single pixel in order to reduce charge resolution below the single-read level (see {\it e.g.} Refs.~\cite{Tiffenberg,Lutz}). This multiple-read strategy allows these CCDs to overcome the 1/f noise issues that otherwise limit the gains of longer integration time. The single-read resolution is achieved by using small, low capacitance read nodes ($\le$ 100~nF) coupled to more conventional amplifiers. The ability to sequentially read pixels allows for scaling this technique to gram-scale detector masses. 

A hybrid design of O(10) charge sensitive segments on a small diamond crystal could achieve sub-electron resolution with a HEMT amplifier design, assuming the charge could be concentrated within the area of a unit cell, and assuming the HEMT design can attain sufficiently small input gate capacitance. Charge transport imaging measurements made in Si demonstrate that at high field strength, both electrons and holes can be focused along charge lines and confined to within 1~mm of the original event location. Thus individual cells can be defined with a maximum size of 1 square millimeter, reducing the capacitance, and therefore resolution, of an individual pixel, at the cost of readout complexity. It is likely that for a dark matter application, this complexity would result in diminishing returns until the segmentation rivaled that of a CCD. It is worth emphasizing that while the development of diamond CCDs is beyond the scope of this paper, there is a lot of synergy between dark matter and many adjacent fields that would all benefit from work towards this end. The crystal dimensions and resolutions of the two proposed detector designs are shown in Table~\ref{tab:detdesigns}.

\subsection{Diamond Calorimetry}
\subsubsection{Phonon Collection Efficiency}

\begin{figure}[t]
    \centering
    \includegraphics[width=3.4in]{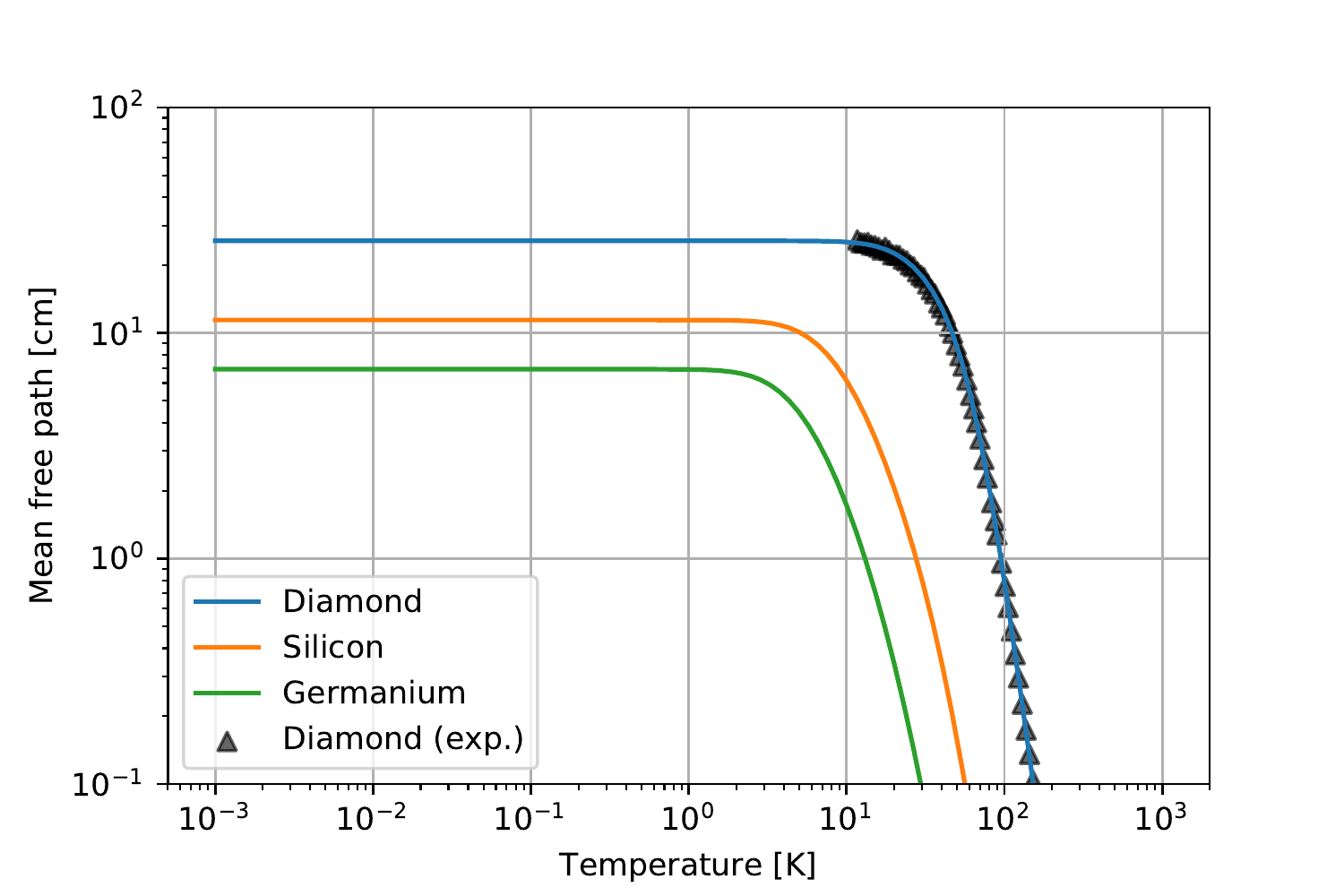}
    \caption{Estimated mean free path using the Callaway model fitted to experimental data \cite{PhysRevB.97.144305,PhysRev.134.A1058} and normalized to the same geometry ($l_b=2.6\; {\rm mm}$) and surface quality ($P=4\times10^{-4}$). The model is described in Ref.~\cite{PhysRevB.97.144305}, with the modification that the specular reflection probability $P$ is a constant rather than a function of phonon frequency.}
    \label{fig:phononmfp}
\end{figure}

Phonon lifetime in diamond can be estimated following the approach in Ref.~\cite{PhysRevB.97.144305}. At low temperatures, the phonon mean free path is completely dominated by boundary scattering and is thus a strong function of crystal dimension and surface quality. In order to compare the phonon mean free path across different materials under the same geometry and surface quality, we fit experimental measurements of thermal conductivity~\cite{PhysRevB.97.144305,PhysRev.134.A1058} to the Callaway model and then normalized parameters related to boundary scattering to the same values across all materials. The results are plotted in Fig.~\ref{fig:phononmfp}. The flattening of the mean free path at low temperatures is a result of the dominance of boundary scattering under which the mean free path is limited by the size of the crystal; for diamond, the crystal is on the order of 1~mm in characteristic size.

To understand the large differences in phonon mean free path shown in Fig.~\ref{fig:phononmfp}, we note that for Ge, Ref.~\cite{tamura} demonstrated that the phonon lifetime is limited by isotopic scattering of phonons. This scattering is due to slight differences in the local potential surrounding Ge atoms of differing atomic weight, which violate lattice symmetry, and allow phonons to decay or scatter. As shown in Table~\ref{tab:properties}, diamond has only two stable isotopes, and natural diamond is about 98.9\% $^{12}\mathrm{C}$, while Si has three stable isotopes, and Ge has four. This results in a much smaller number of isotopic scattering sites for typical diamond crystals, as compared to Si and Ge. In diamond crystals which are isotopically enriched by a factor of 15, for example, thermal conductivity, and thus phonon mean free path, were increased by a factor of 2~\cite{Neves2001}. 

We note that under the same geometry and surface quality, the phonon mean free path in diamond is still several times higher than that in Si and Ge. The phonon lifetime $\tau_{\rm life}$ can be estimated from the mean free path $\ell$ via a simple conversion:
\begin{equation}
    \tau_{\rm life} = \ell / c_s
\end{equation}
where $c_s$ is the average speed of sound. Both values are listed in Table~\ref{tab:properties} for diamond, Si, and Ge. At low temperatures, this gives a phonon life time of $\sim$20~$\mathrm{\mu sec}$ in diamond for a mm-sized crystal.

The phonons produced need to be read out via some phonon sensor. Here we use as a point of reference the quasiparticle-trap-assisted electrothermal-feedback transition edge sensor (QET), which uses aluminum phonon-absorbing fins to channel phonon energy into a highly sensitive temperature to current sensor, the transition edge sensor (TES)~\cite{Romani,kurinskyThesis}. The collection efficiency across the substrate-absorber interface has to be high to maintain good energy resolution. A good benchmark is to aim for phonon collection times below 1~$\mu$sec, as typical low-$T_c$ TESs typically have response times of a few $\mu$sec. 

Let us estimate this efficiency for a diamond substrate with an aluminum/tungsten (Al/W) QET like that described in Ref.~\cite{kurinskyThesis}. The time constant for phonon collection at the interface is given by~\cite{Hochberg:2015fth}
\begin{equation}
    \tau_{\rm collection} = \frac{4V_{\rm crystal}}{A_W c_s f_{\rm W}+A_{\rm Al}c_s f_{\rm Al}}
\end{equation}
where $V_{\rm crystal}$ is the volume of the crystal, $A_{\rm W,Al}$ is the surface area of the absorber (tungsten or aluminum), $c_s$ is the average speed of sound and $f_{\rm W,Al}$ is the transmission probability across the crystal-tungsten or crystal-aluminum interface. The transmission probability can be estimated using the acoustic mismatch model described in Ref.~\cite{Kaplan:1979}. The corresponding values for different materials are tabulated in Table~\ref{tab:properties}. We note that the high speed of sound in diamond leads to a short collection time.

\begin{table*}[t]
    \centering
    \begin{tabular}{|l|c|c|c|c|c|c|c|}
    \hline
    Readout & Design & Dimensions & Mass (mg) & Temp. (K) & $V_{\rm Bias}$ & $\sigma_E$ & $\sigma_{q}$ \\
    \hline
    \multirow{2}{*}{Charge} & Single Cell & $16~{\rm mm}^2\times 0.5~{\rm mm}$ & 28 & 4.2~K & 10~V & 13--39~eVee & 1--3$e^{-}$ \\
    & Segmented & $1~{\rm mm}^2\times 0.5~{\rm mm}$ & 1.8 & & & 1.3--3.9~eVee & 0.1--0.3$e^{-}$/segment \\
    \hline
    & A & $16~{\rm mm}^2\times 0.5~{\rm mm}$ & 28 & 40~mK & 40~V & 0.2~eV & $\SI{5e-2}e^{-}$ \\
    Phonon & B & $100~{\rm mm}^2\times 0.5~{\rm mm}$ & 275 & 20~mK & 30~V & 30~meV & $\SI{1e-3}e^{-}$ \\
     & C & $16~{\rm mm}^2\times 0.5~{\rm mm}$ & 28 & 10~mK & 0~V & 2~meV & - \\
    \hline
    \end{tabular}
    \caption{Summary of the detector designs discussion. Voltage bias for the charge designs should be high enough to ensure full charge collection, and need not be higher. For the phonon designs, the voltage sets the charge resolution, but for low enough resolution the primary region of interest is below the bandgap, meaning no applied voltage is necessary. Crystal sizes are by those currently available combined with the assumption that cm-scale crystals are attainable.}
    \label{tab:detdesigns}
\end{table*}

The overall collection efficiency and phonon pulse time can be obtained by combining phonon collection time with phonon lifetime as \cite{Hochberg:2015fth}:
\begin{equation}
    \tau_{\rm pulse}^{-1} = \tau_{\rm life}^{-1} + \tau_{\rm collection}^{-1}
\end{equation}
where, as a result,
\begin{equation}
    f_{\rm collection} = \frac{\tau_{\rm life}}{\tau_{\rm life}+\tau_{\rm collection}}
\end{equation}
To determine the overall energy efficiency of the QET we also need to include the phonon to quasiparticle conversion efficiency $f_{\rm conversion}$ in the aluminum fin as well as quasiparticle collection efficiency $f_{\rm qp}$ at the Al-W interface. From past experience with CDMS detectors, these efficiencies are $\sim60\%$ and $\sim75\%$ respectively \cite{kurinskyThesis}. Thus the overall energy efficiency is given by:
\begin{align}
    \epsilon &= f_{\rm collection}f_{\rm conversion}f_{\rm qp} \\
             &\approx 0.6\times 0.75 \times f_{\rm collection}
\end{align}
For a mm-sized diamond crystal with 70\% aluminum coverage, the overall efficiency is estimated to be around 44\%. 

For the resolution estimates in our reference designs, we take this to be an upper limit, and assume crystal purity and sensor non-idealities reduce the efficiency somewhat. As a means of contextualizing this number, efficiencies in excess of 20\% have been achieved in large Si and Ge detectors, and an absolute limit of $\sim$60\% is expected due to the phonon down-conversion process~\cite{kurinskyThesis}.

\subsubsection{Energy Resolution}

\begin{table*}[th]
    \centering
    \begin{tabular}{|c|c|c|c|c|}
        \hline
        \multicolumn{2}{|c|}{} & \multicolumn{3}{c|}{Design} \\
        \hline
        Parameter & Description & A & B & C  \\
        \hline
        $\gamma$ & Specific Heat & \multicolumn{3}{c|}{108 $\mathrm{J m^{-3}K^{-1}}$} \\
        $\mathcal{L}-1$ & Transition Sharpness & \multicolumn{3}{c|}{30} \\
        $\eta_{\rm TES}$ & TES Thickness & \multicolumn{3}{c|}{40~nm} \\
        $N_{\rm TES}$ & Number of TES in Array & \multicolumn{3}{c|}{100} \\
        $\xi_{\rm Al}$ & Aluminum coverage ($A_{\rm Al}/w_{\rm crystal}l_{\rm crystal}$) & \multicolumn{3}{c|}{70\%} \\
        $\tau_{\mathrm{life}}$ & Phonon Lifetime & \multicolumn{3}{c|}{19.2~$\mathrm{\mu s}$} \\
        $\tau_{\mathrm{collect}}$ & Phonon Collection Time & \multicolumn{3}{c|}{335 ns} \\
        $\tau_{\mathrm{pulse}}$ & Phonon Pulse Time & \multicolumn{3}{c|}{330 ns} \\
        $f_{\mathrm{collect}}$ & Phonon Collection Efficiency & \multicolumn{3}{c|}{98.3\%} \\
        $\epsilon$ (ideal)  & Ideal Energy Efficiency & \multicolumn{3}{c|}{44\%} \\
        $w_{\rm TES}$ & TES Width & 2.4~$\mathrm{\mu m}$ & 1.2~$\mu$m & $400$~nm \\
        $l_{\rm TES}$ & TES Length & 50 $\mu$m & 25 $\mu$m & 10 $\mu$m\\
        $T_c$ & Critical Temperature & 60~mK & 40~mK & 15~mK \\
        $T_b$ & Crystal Temperature & $\le$30~mK & $\le$20~mK & $\le$8~mK \\
        $\tau_{\rm TES}$ & TES Response Time & 20~$\mu$s  & 70~$\mu$s & 1.3~ms \\
        $\epsilon$ & Design Energy Efficiency & 10\% & 20\% & 25\% \\
        $\sigma_{\rm E}$ & Energy Resolution & 200~meV & 30~meV & 2~meV \\
        \hline
    \end{tabular}
    \caption{Calorimetric detector parameters used for resolution calculations. }
    \label{tab:tes}
\end{table*}

The main limitation of a charge readout for diamond is its larger bandgap compared to many materials already employed by other direct detection searches. The real strength of diamond as a future detector material is, however, the combination of high carrier mobility and long-lived, high energy phonon excitations.

To see how these qualities factor into the energy resolution of the device, consider the energy resolution of a diamond calorimeter with a TES~\cite{Irwin} readout as part of a QET array~\cite{kurinskyThesis}, as described earlier. The minimum resolution (limited only by thermal fluctuation noise) for a given detector design obeys the relation~\cite{kurinskyThesis}
\begin{equation}
\sigma_e = \sqrt{\frac{2k_b T_c^2 G}{\epsilon^2}\left[\tau_{\rm pulse}+\frac{2}{5}\tau_{\rm TES}\right]},
\end{equation}
where $T_c$ is the TES critical temperature, $G$ is the thermal conductance of the TES to the crystal, $\epsilon$ is the phonon collection efficiency, $\tau_{\rm pulse}$ is the decay time of the phonon pulse, and $\tau_{\rm TES}$ is the TES response time in electrothermal feedback. This last term depends on both the thermal conductance $G$ and heat capacity $C$ of the TES as well as the electrothermal closed-loop gain $\mathcal{L}$ (which is directly related to the slope of the TES transition curve) as
\begin{equation}
\tau_{\rm TES} = \frac{C}{G}\frac{1}{\mathcal{L}-1}.
\end{equation}
This means that for an arbitrarily fast collection time, we have a minimum QET resolution of
\begin{equation}\label{eq:res}
\sigma_{e} \geq \frac{\sqrt{4k_b T_c^2 C}}{\sqrt{5}\epsilon} \approx \frac{1}{\epsilon}\sqrt{\frac{2k_b \gamma T_c^3 V_{\rm TES}}{(\mathcal{L}-1)}}
\end{equation}
where $\gamma$ is the specific heat of the TES in the normal state, and $V_{\rm TES}=N_{\rm TES}l_{TES}w_{\rm TES}\eta_{\rm TES}$ is the volume of the sensor, with dimensions described in Table~\ref{tab:tes}. An additional factor of 2.4 is due to the jump in heat capacity at the superconducting transition predicted by BCS superconducting theory~\cite{Irwin}. The clear trade-off for energy resolution is thus between critical temperature and sensor volume, and the transition becomes sharper for a TES with bulk-like critical temperature. Using magnetic impurities to alter $T_c$ has been shown to work well~\cite{youngTc,youngTc2,Roth}, but this doping process can reduce the sharpness of the transition by creating a variance in $T_c$ throughout the film, so it is preferable to use films as close to intrinsic $T_c$ as possible.

Over the past two decades, significant progress has been made producing detectors with sub-eV resolution using TESs made of tungsten~\cite{cabrera}, as well as molybdenum and titanium bilayers (see {\it e.g.} Refs.~\cite{AlTi,TiBilayer,TESMultiplexing}). The success of these materials comes primarily from their small electron-phonon coupling and low critical temperatures, which allow for low thermal fluctuation noise, on the order of $1~\mathrm{aW/Hz}^{1/2}$~\cite{fink}. This low thermal conductance is limited by the small electron-phonon coupling within the tungsten, and not by the interface with the absorber, resulting in a device performance largely independent of the properties of the coupled absorber~\cite{Irwin,kurinskyThesis,AlTi}. It is thus straightforward for us to apply scalings to diamond based on performance of tungsten QETs on Si and Ge, with the only difference being the interaction of the phonons with the QET absorber. These are easily extensible to other TES materials in terms of thermal performance, though the efficiency of transport from the Al fins to the TES for these materials is hard to estimate.

\begin{figure}[t]
    \centering
    \includegraphics[width=3.4in]{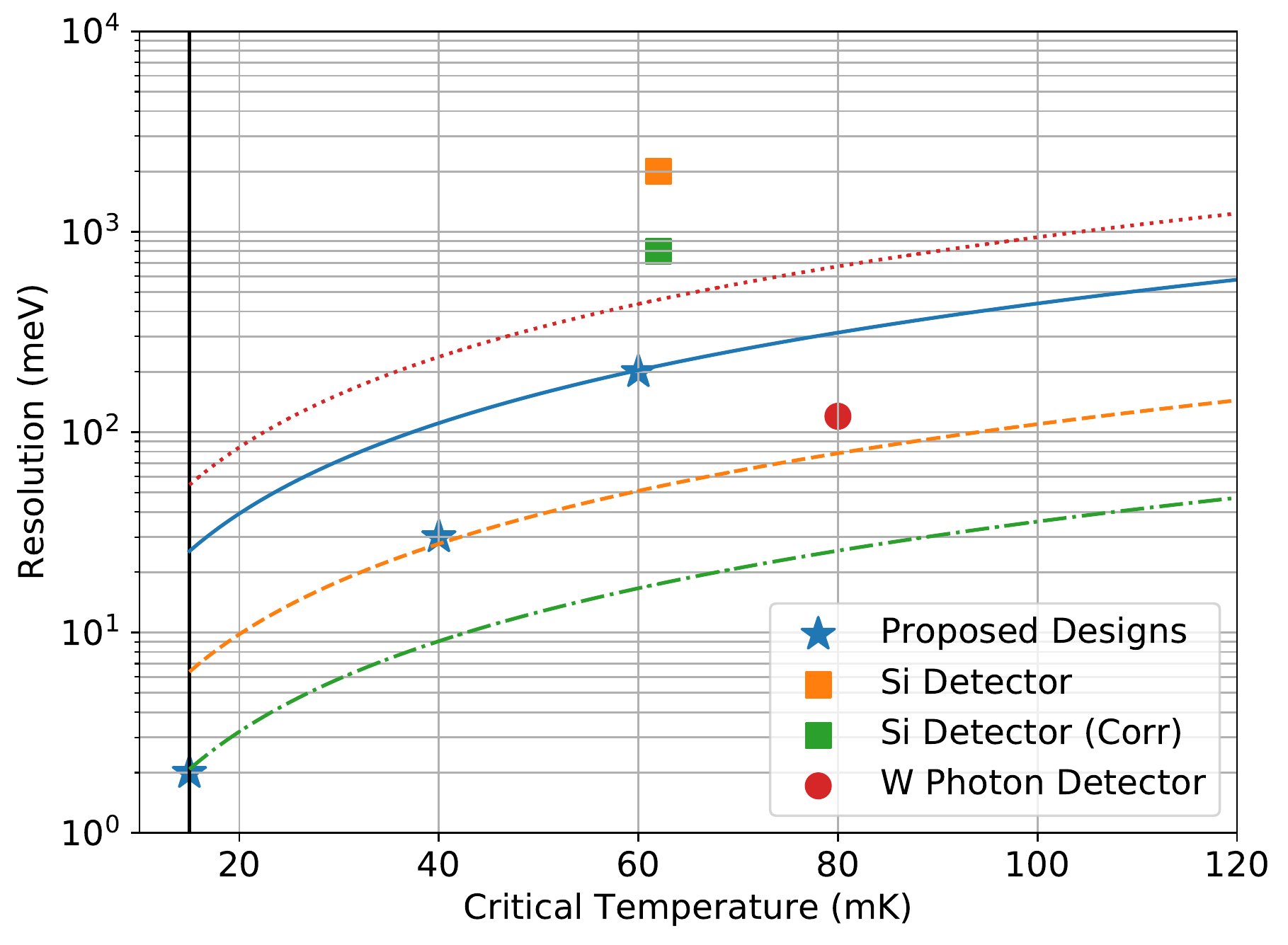}
    \caption{Energy resolution scaled according to Eq.~\eqref{eq:res} using the parameters in Table~\ref{tab:tes}. These are shown along with the scaling relations for the given volume and efficiency as a function of $T_c$ (solid line for design A, dashed for design B, and dot-dashed for design C). Also shown are the resolution demonstrated in a tungsten TES photon detector~\cite{cabrera} and a Si detector with tungsten QET readout~\cite{hong}. The corrected resolution of the Si detector for upgraded readout electronics is also shown, compared to its ideal scaling relation (dotted red line).}
    \label{fig:resScaling}
\end{figure}

Table~\ref{tab:tes} gives the parameters for three progressively more aggressive diamond detector designs targeting meV-eV scale resolutions, assuming Al/W QETs. The scalings employed to compute the expected resolution are shown in Fig.~\ref{fig:resScaling}, in comparison with other resolution measurements. Design A represents the implementation of demonstrated device performance on a Si 1~cm$^2\times$~1~mm crystal optimized for the smaller diamond crystal, and assuming upgrades to the SuperCDMS SNOLAB cold electronics already demonstrated on TES test structures~\cite{kurinsky,kurinskyThesis}. To be conservative, design A also assumes a lower efficiency than has been achieved in QETs on Si and Ge (20--25\% for ideal devices~\cite{kurinsky}) due to the phonon impedance mismatch and the potential for increased boundary scattering losses. Designs B and C assume the same readout electronics, but depend on improvements in the fabrication process to achieve progressively smaller TES linewidth in photolithography, and that the efficiency of the QET can be improved for smaller TESs without increasing the effective heat capacity or thermal conductance of the TES. 

Design C is very aggressive, and we assume there is no additional resolution limitation from phonon quantization; this is approaching the limit of this technique given that bulk tungsten is expected to have a $T_c$ of 15~mK, and the TES volume employed is likely approaching a lower volume limit. This lower volume limit is set by limitations on the inductance and resistance of SQUID input coils used as the first-stage current amplifiers, and the minimum feature size necessary to achieve a stable (and operable) $T_c$ in tungsten films. An alternative way to achieve meV-scale resolution is by employing a TES multiplexing scheme (see e.g. Ref.~\cite{TESMultiplexing}) in which each small TES has a low resolution, and employing the position dependence of the signal to reduce the amount of TES area necessary to integrate. 

As an aside on the optimal QET design for a given target material, the phonon energy scale determines the maximum $T_c$ of the superconducting absorber as
\begin{equation}
E_g = \frac{7}{2}k_bT_c.
\end{equation}
For example, for Si and Ge, a mean athermal acoustic phonon energy of around 3~meV~\cite{SundqvistThesis} means that Nb (with a $T_c$ of $\sim$9K, and an energy gap of 3~meV) will have a much lower collection efficiency than Al, which has an energy gap of $\sim$0.3~meV and a $T_c$ of $\sim$1K~\cite{Irwin}. The energy scale of the acoustic phonons scales with the sound-speed $c_s$, and phonon momentum $q$, as $E_{ac}=\hbar q c_s$~\cite{Jacoboni,SundqvistThesis,kurinskyThesis}. The higher drift momenta (higher $q$) and larger sound speed in diamond suggest a mean phonon energy in excess of 10~meV even at moderate field strength, meaning high-purity superconducting films of $T_c$ up to $\sim$10~K may be employed as efficient phonon absorbers.

In addition, as previously mentioned, the phonon collection time and efficiency are greatly influenced by the boundary impedance at the target/absorber interface. This impedance is larger for mismatched atomic mass, and becomes transparent as the atomic weight of the absorber and target approach parity. An interesting prospect would be the use of superconducting diamond (produced by heavily doping diamond thin films with Boron~\cite{SCD1,SCD2}) as the QET absorber. Superconducting diamond thin films show tunable $T_c$ in the range 4--11~K~\cite{SCD2}. If long quasiparticle lifetime can be demonstrated in these films, the possibility of achieving down-conversion-limited QETs with efficiency exceeding 50\%, and faster fall-times, would allow diamond QETs to achieve their ultimate bandwidth and efficiency limits.

\section{Backgrounds}\label{sec:backgrounds}

While we do not include a background model in the sensitivity projections we present below, it is nevertheless prudent for us to include a discussion of possible backgrounds in this first paper. We leave a more detailed background study to a future work, due to the dependence of such a study on details of the experimental setup. For a more in-depth discussion of backgrounds relevant to low-mass dark matter searches, we refer the reader to Refs.~\cite{CDMSSensitivity,Agnese:2018col,hertel}.

\subsection{Excitation of Sub-Gap States}
The largest internal background in terms of the raw event rate that affects single-charge and photon counting devices is the `dark rate'---the random occurrence of single and multiple charge events due to the decay or excitation of low energy states. The three current electron-recoil dark matter searches of Refs.~\cite{Essig:2012yx,Agnese:2018col,sensei} were all background-limited only due to such dark rates at low mass.

For CCDs, there is a well-established correlation between operating temperature and dark rate, indicating that shallow impurity sites and crystal defects are the dominant contribution to this leakage. Smaller bandgap semiconductors also tend to have shallower impurity wells, making them more susceptible to thermal carrier generation. Almost all shallow level impurities in a given semiconductor will have binding energies of the same order of magnitude as a hydrogenic impurity. Thermal excitation of charge from impurity states of energy $E_{T}$, with number density $n_{I}$, by a low-temperature ($\lesssim10$K) blackbody of temperature $T$, yields an event rate $\Gamma$ that obeys the proportionality~\cite{kurinsky}\footnote{For higher temperature blackbody radiation, this is still a good approximation but is no longer exact.}
\begin{equation}
\Gamma(n_{I},E_{T},T) \propto n_{I}T^4E_{T}^3\exp{\left(\frac{E_T}{k_bT}\right)}\,.
\end{equation}
Moving to a semiconductor with a larger bandgap and deeper impurity wells thus reduces this dark rate substantially, at the expense of a larger energy threshold. The characteristic binding energies for several impurities in Si and diamond are given in Table~\ref{tab:impurity}. Also included are the binding energies of neutral impurities, in which an extra charge may be bound to an otherwise neutral impurity atom. These are only stable at very low temperatures, but are important for cryogenic detectors.

To give some concrete numbers, the DAMIC collaboration recently achieved a dark rate of $4e^{-}\mathrm{mm^{-2}d^{-1}}$, using a science-grade CCD fabricated on a high-purity silicon substrate cooled to $\sim$105~K~\cite{damicDP}. Converting this to more conventional units, for a 675~$\mu$m active depth, this corresponds to a leakage rate of $\sim 6e^{-}\mathrm{mm^{-3}d^{-1}}$ in Si, or 30~mHz/g.

Given the same impurity density and temperature, if we assume this is driven by {\it e.g.} lithium impurities, the change in binding energy gives a rate reduction of three orders of magnitude in leakage. The much lower leakage current in diamond electronics than comparable Si electronics is evidence of this effect at work~\cite{Neves2001}. It is therefore possible that at cryogenic temperatures for sufficiently pure diamond substrates, leakage rates less than $10^{-4} \mathrm{Hz/g}$ ($10^3 e^{-}\mathrm{g^{-1}yr^{-1}}$) may be achievable.

\begin{table}
\centering
\begin{tabular}{|l|c|c|c|c|}
\hline
\multirow{2}{*}{Type} & \multirow{2}{*}{Element} & \multicolumn{3}{c|}{Binding Energy (meV)} \\
 & & Diamond & Si & Ge\\
\hline
\multirow{3}{*}{Donor} & N & 1700, 4000 & 15--50 & - \\
& P & 500 & 45 & 12 \\ 
& Li & 230 & 33 & 9.3 \\ 
\hline
Acceptor & B & 370 & 45 & 10 \\ 
\hline
Neutral & - & $\sim$10 & 2 & 0.5 \\
\hline
\end{tabular}
\caption{Energies of common residual impurities in diamond, Si and Ge in units of meV~\cite{Neves2001,ridley}. Given the difficulty to controllably dope Si and Ge with nitrogen, the impurity energy is not well-determined for Si and essentially unmeasured for Ge, though it should be on the order of the other shallow impurities.}
\label{tab:impurity}
\end{table}

\subsection{Electronic Recoils}

The largest background in most dark matter searches at higher masses stems from electron recoils from minimum ionizing radiation and radioactivity in the laboratory. While most minimum ionizing particles will deposit energies much larger than the energy range of interest, high-energy photons can deposit small energy deposits due to Compton recoils. For a given background, the rate of these recoils depends on the photon cross-section in diamond as compared to Si and Ge, which in turn is proportional to the electron density. The higher-energy bandgap of diamond limits the energy of these recoils to $\gtrsim$5.5 eV, meaning that the nuclear recoil space below this energy is by definition free of electron-recoils. In addition, the fact that diamond has fewer total electrons per lattice site leads to fewer scattering sites per nucleon, and thus a smaller photon cross-section per gram. For these reasons, a diamond detector, placed in the shielding of an existing experiment, will always be less susceptible to high-energy photons than heavier elements in the energy region of interest for low-mass DM searches.

For nuclear-recoil searches for dark matter, the cryogenic detector designs discussed in the previous section benefit from the larger bandgap in diamond due to the fact that a small voltage across the crystal will ensure that all electron recoils, which produce at least one electron-hole pair, can be boosted outside of the energy range of interest. In contrast, nuclear recoils at this energy scale ($<100\;{\rm eV}$) have very low charge yield, and will be predominantly zero-charge events. This is thus true of any detector with an energy resolution much less than the energy of its charge quanta. In addition, the ability of these detectors to highly resolve electron-hole pairs in larger field strengths, or by varying the field strength, allows for reconstruction of charge yield and recoil energy. This means that electron recoils can be rejected in an analogous way to the two-phase detectors currently in use.

\subsubsection*{Cosmogenic Backgrounds}

Following the example of Ref.~\cite{CDMSSensitivity}, we recognize that the ultimate sensitivity limit of this technique will depend on the radiopurity of the crystal. As with Si and Ge, cosmogenically produced tritium will be an irreducible background, though the exact content will depend on crystal history and initial composition. 

$\mathrm{^{14}C}$, on the other hand, is likely to present an acute challenge. This isotope decays by $\beta^{-}$ with a broad energy distribution with a mean energy of 48~keV and a maximum energy at 160~keV~\cite{C14}, with a half life of about 5700 years. The natural abundance of $\mathrm{^{14}C}$, around 1 part per trillion ($\mathrm{^{14}C/C}\sim10^{-12}$)~\cite{c14concentration}, therefore implies an event rate of around 0.2 Hz/g. The vast majority of these events will be outside the energy range of interest; we only care about events with energies below about 100~eV, which will occur at a rate of about 100 $\mu$Hz/g, equivalent to about 10 events per gram-day. Thus, without any ability to reject these events, this concentration of $\mathrm{^{14}C}$ would limit diamond detectors to gram-day exposures.

The $\mathrm{^{14}C}$ concentration can be greatly reduced, however, by employing sources of carbon with a large overburden which prevents cosmogenic production of $\mathrm{^{14}C}$, which therefore have a much smaller abundance of the radioisotope. Underground sources of carbon, such as methane deposits, have been shown to have natural abundances of $\mathrm{^{14}C/C}\sim10^{-18}$, and mass spectroscopy methods promise to reduce the $\mathrm{^{14}C/C}$ ratio even further, to $\mathrm{^{14}C/C}\lesssim10^{-21}$~\cite{c14concentration}. These improved abundances would correspond to only about 10 events per kg-year and 1 event for 100 kg-years respectively, which allow for background-free searches for the exposures considered in this paper. Given that the seed material for CVD diamond can be very precisely controlled, and that impurities are removed in a subsequent refinement step, it is likely that CVD diamonds with much lower than natural abundance of $\mathrm{^{14}C}$ can be made without much additional effort. Careful selection of seed material will thus be important to obtaining low activation diamonds.

For arrays of small crystals, or a segmented charge design, it is possible that the electron track from a $\mathrm{^{14}C}$ decay may extend between multiple detectors, allowing for a multiples veto, but some fraction of these electrons will only deposit a small amount of energy. These events will always generate at least the initial charge, so they will not be a background for the NR searches proposed here below the bandgap energy, but will be the ultimate limit for the dark photon and electron-recoil dark matter searches. The expected $\mathrm{^{14}C}$ abundance therefore helps determine the optimal size of each diamond detector, in that the rate must be low enough to reduce dead time from high-energy events, and we would like to maximize the probability of observing a multiple scatter from low-energy $\mathrm{^{14}C}$ events.

\subsection{Nuclear Recoils}

The one area in which backgrounds may be worse for a diamond detector are for nuclear recoils, in particular neutron backgrounds. The smaller atomic weight of carbon, which makes diamond more attractive for low-mass dark matter, also allows these particles to more efficiently transfer momentum by nuclear recoils, resulting in a background that extends to higher energies for a given neutron flux and energy. In addition, the neutron scattering cross-section for thermal neutrons on carbon atoms is about 2.5 times higher than for silicon atoms~\cite{neutronScattering}. This means that there will be both a higher rate and larger energy dispersion, making neutrons a more insidious background for carbon-based detectors. Helium, another candidate material for light dark matter detection, has a 4 times lower cross section, despite having a higher energy transfer efficiency~\cite{hertel}.

An additional consideration, as with other DM search media, is the impact degraded alphas will have on the background spectrum for diamond detectors. These alphas are not necessarily intrinsic to diamond, but instead originate from high energy decays in the detector housing which produce an alpha particle. If moderated, this alpha particle can deposit a small amount of energy at the surface of a detector and, if not fiducialized, could lead to a smooth, DM-like background which is hard to reject. This is distinctly different than the neutron background, which is irreducible and isotropic, but this background is unlikely to be rejectable without full fiducialization ability. Both designs are, in principle, capable of some degree of position reconstruction in three dimensions, either taking advantage of anisotropic charge propagation or phonon diffusion~\cite{kurinskyThesis}, but a more quantitative statement would depend on the specific design chosen. It is, on the other hand, possible to reduce this background by careful design of the detector housing with high-purity copper and similar materials as is being done for the next generation of DM experiments (see {\it e.g.} Ref.~\cite{CDMSSensitivity}).

\section{Sensitivity Projections}\label{sec:results}

In this section we discuss the reach of diamond detectors for absorption of bosonic dark matter (\ref{sec:abs}), detection of electron-recoiling dark matter (\ref{sec:erdm}), and nuclear-recoil dark matter (\ref{sec:nr}). We do not necessarily assume a particular detector design out of those described in this paper, but instead rely on them as a proof of principle that the detection thresholds we employ in our projections are realistic. The exception is for the purely calorimetric measurement described in Section~\ref{sec:nr}, in which the exact threshold will impact the physics reach of the detector. In all other cases we assume sub-electron thresholds, in which case the actual threshold value is not important in the background-free limit. We will consider projections with background-free exposures from 1~gram-day to 1~kg-year, corresponding to the exposure possible with a first-generation R\&D detector up to a large-scale diamond-based experiment.

For all of our reach calculations, the local dark matter density $\rho_{\chi}$ is taken to be $0.3\; {\rm GeV}/{\rm cm}^3$; the dark matter is assumed to follow a Maxwell-Boltzmann velocity distribution~\cite{billard} with mean velocity of 220 km/sec and escape velocity of 500 km/sec. All our projections are 95\% C.L. which corresponds to 3 signal events.

\begin{figure}[t!]
\begin{center}
\includegraphics[width=0.48\textwidth]{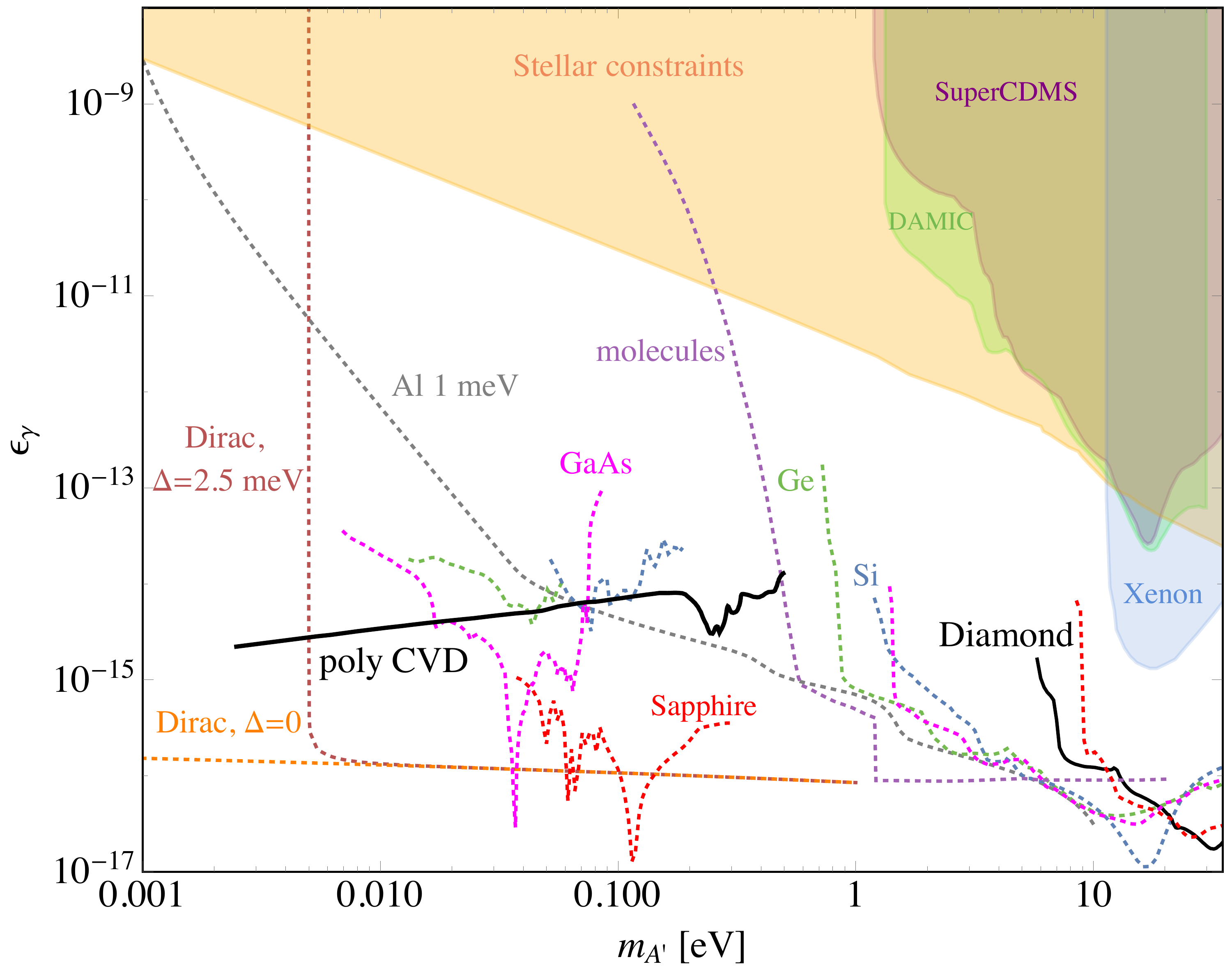}
\caption{ \label{fig:absphoton}
Projected reach at 95\% C.L. for absorption of kinetically mixed dark photons with mass $>1$~meV. The solid black curves indicate the expected reach for a kg-year exposure of diamond. Projected reach for germanium and silicon~\cite{Hochberg:2016sqx}, Dirac materials~\cite{Hochberg:2017wce}, polar crystals~\cite{Griffin:2018bjn}, molecules~\cite{Arvanitaki:2017nhi} and superconducting aluminum~\cite{Hochberg:2016ajh} targets are indicated by the dotted curves. Constraints from stellar emission~\cite{An:2013yua,An:2014twa}, DAMIC~\cite{Aguilar-Arevalo:2016zop}, SuperCDMS~\cite{Agnese:2018col} and Xenon~\cite{An:2014twa} data are shown by the shaded orange, green, purple and blue regions, respectively.}
\end{center}
\end{figure}

\subsection{Dark Matter Absorption}\label{sec:abs}

We begin with the potential of diamond detectors to probe bosonic dark matter via an absorption process. The rate for DM absorption (in counts per unit time per unit mass) is given by
\beq
R_{\rm abs}=\frac{1}{\rho_T} \frac{\rho_{\rm \chi}}{m_{\rm \chi}}\langle n_T \sigma_{\rm abs} v_{\rm rel}\rangle_{\rm DM}\,,
\eeq
where $\rho_T$ is the target mass density, $\rho_\chi$ is the DM mass density, $m_\chi$ is the DM mass, $n_T$ is the number density of the target, $\sigma_{\rm abs}$ is the absorption cross section, and $v_{\rm rel}$ is the relative velocity between the DM and the target. 

By relating the absorption cross section of DM to that of photons, it is possible to translate measurements of optical conductivity in the material into the projected reach for DM absorption. For photons, the optical conductivity is related to the absorption cross section via the optical theorem: 
\beq\label{eq:abs}
\langle n_T \sigma_{\rm abs} v_{\rm rel}\rangle_\gamma = -\frac{{\rm Im}\,\Pi(\omega)}{\omega}=\omega\; {\rm Im}\epsilon_r\,,
\eeq
Here $\omega$ is the energy of the incoming photon, and $\Pi(\omega)$ the in-medium polarization tensor in the relevant limit of $|{\bf q}| \ll \omega$. For DM absorption, the incoming energy is $\omega\sim m_\chi$ and the incoming momenta $|{\bf q}|\sim 10^{-3} m_\chi$, such that the longitudinal and transverse parts of the polarization tensor are roughly equal and $\Pi_L \approx \Pi_T \equiv \Pi=\omega^2(1-\epsilon_r)$, with $\epsilon_r$ the complex permittivity, related to the complex index of refraction $\tilde n$ via $\epsilon_r = \tilde n^2 = (n+ik)^2$. The sensitivity of the material to absorption of DM is thus obtained by relating the absorption process of DM to that of photons in the material through the permittivity.

{\bf Dark photons.} For a kinetically mixed dark photon $A'$, with ${\cal L}\supset -\frac{\epsilon_\gamma}{2} F_{\mu\nu} F'^{\mu\nu}$, the effective mixing angle between the photon and dark photon is medium-dependent, and is given by
\beq\label{eq:kappaeff}
\epsilon^2_{\gamma,\rm eff} =  \frac{ \epsilon_\gamma^2 m_{A'}^4}{ \left[ m_{A'}^2 - {\rm Re}\;\Pi(m_{A'}) \right]^2 + \left[ {\rm Im}\;\Pi(m_{A'}) \right]^2} \,,
\eeq
with $m_A'$ the mass of the dark photon.
The rate of absorption is then
\beq
	R_{\rm abs}^{A'}=  \frac{1}{\rho_T}  \rho_\chi \varepsilon_{\gamma,\rm eff}^2 \,  {\rm Im}\,\epsilon_r \,.
\eeq
We use measurements of the complex index of refraction in carbon from Refs.~\cite{refindex1, refindex2}, including above-gap and sub-gap processes, in similar spirit to Ref.~\cite{Hochberg:2016sqx}. We assume the dark photons to comprise the entirety of local dark matter density. Our results are presented in Fig.~\ref{fig:absphoton}. The solid black curves indicate the 95\% C.L. expected reach in diamond for a kg-year exposure, corresponding to 3 events, via electronic and sub-gap phonon excitations. For comparison, also shown in the dotted curves are the projected reach of superconducting aluminum targets~\cite{Hochberg:2016ajh}, semiconductors such as silicon and germanium~\cite{Hochberg:2016sqx}, Dirac materials~\cite{Hochberg:2017wce}, polar crystals~\cite{Griffin:2018bjn} and molecules~\cite{Arvanitaki:2017nhi}. Stellar emission constraints~\cite{An:2013yua,An:2014twa} are shown in shaded orange, while the bounds from DAMIC~\cite{Aguilar-Arevalo:2016zop}, SuperCDMS~\cite{Agnese:2018col} and Xenon data~\cite{An:2014twa} are shown in shaded green, purple and blue, respectively. 

\begin{figure}[t!]
\begin{center}
\includegraphics[width=0.48\textwidth]{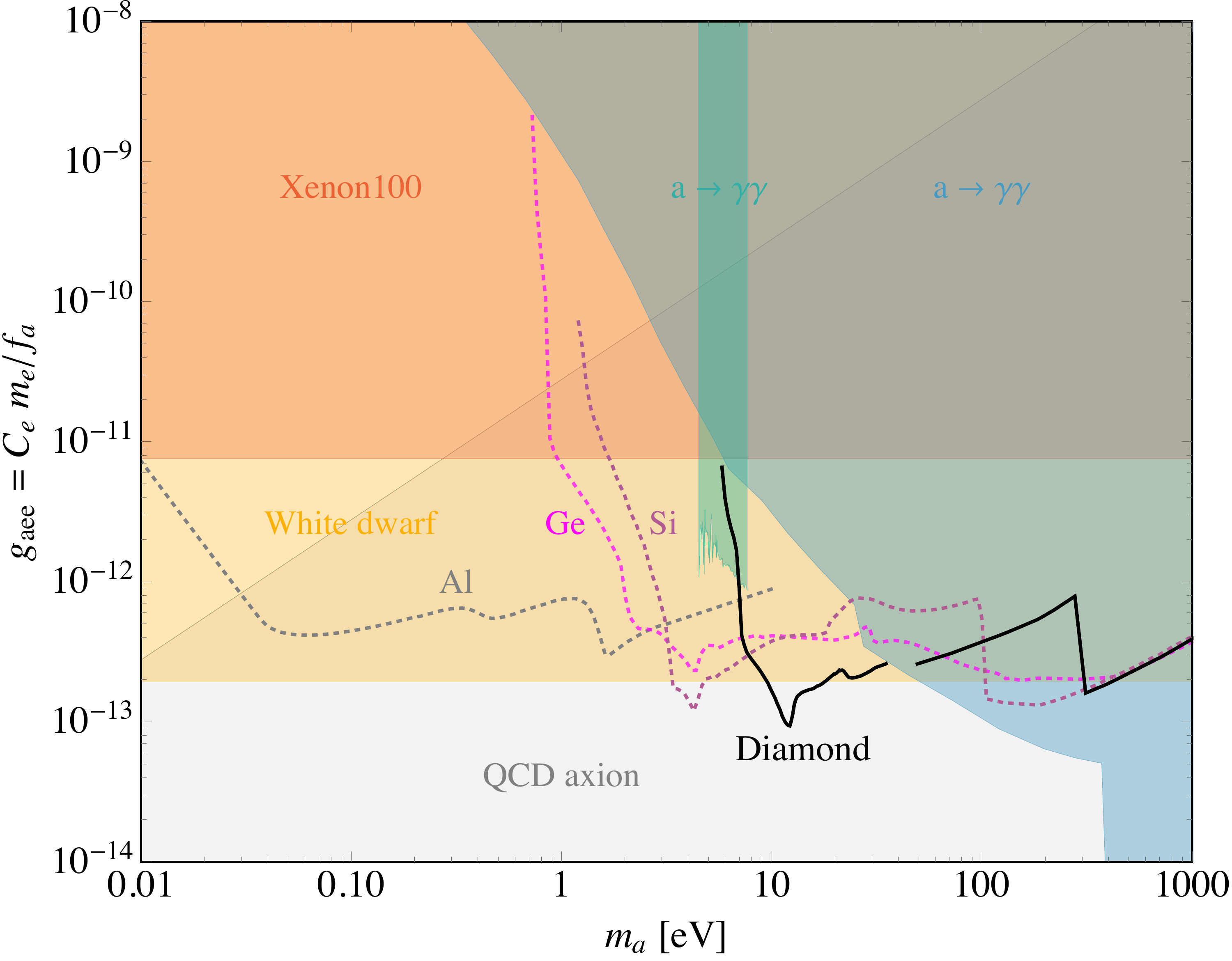}
\caption{ \label{fig:absALP}
Projected reach at 95\% C.L. for absorption of axion-like particles. The reach of a kg-year exposure of diamond is shown by the solid black curve.  The reach for semiconductors such as germanium and silicon~\cite{Hochberg:2016sqx} and superconducting aluminum~\cite{Hochberg:2016ajh} targets is depicted by the dotted magenta, purple and gray curves, respectively. Stellar constraints from Xenon100 data~\cite{Aprile:2014eoa} and white dwarfs~\cite{Raffelt:2006cw}  are indicated by the shaded red and orange regions, respectively. Constraints from loop-induced couplings to photons are presented in the shaded blue regions~\cite{Grin:2006aw,Arias:2012az}. The QCD axion region of interest is indicated in shaded gray.}
\end{center}
\end{figure}

{\bf Axion-like particles.} 
For an axion-like particle $a$ with mass $m_a$ that couples to electrons, 
\begin{equation}
	{\cal L}\supset \frac{g_{aee}}{2 m_e} (\partial_\mu a)\bar e \gamma^\mu \gamma^5 e\,,
\end{equation}
the absorption rate on electrons can be related to the absorption of photons via the axioelectric effect, and is given by
\begin{equation}
\label{eq:rateAxion}
	R_{\rm abs}^a = \frac{1}{\rho_T} \rho_\chi  \frac{3 m_a^2}{4 m_e^2}  \frac{g_{aee}^2}{e^2}\, {\rm Im}\,\epsilon_r\,.
\end{equation}
We use the measurement of Ref.~\cite{refindex1} of electronic excitations, together with the semi-analytical theoretical computations of Henke et al~\cite{refindex3} for carbon. 
The resulting projected reach of a diamond detector on the absorption parameter space of axion-like particles for a kg-year exposure is shown in Fig.~\ref{fig:absALP} by the solid black curves. For comparison, the reach of superconducting aluminum~\cite{Hochberg:2016ajh} targets as well as silicon and germanium~\cite{Hochberg:2016sqx} is also shown by the dotted curves. Constraints from white dwarfs~\cite{Raffelt:2006cw} and Xenon100 data~\cite{Aprile:2014eoa} are indicated in the shaded orange and red regions, respectively. Constraints arising from the model-dependent loop-induced couplings to photons are shown in shaded blue~\cite{Grin:2006aw,Arias:2012az}. The QCD axion region of interest is depicted by the shaded gray area. As is evident, diamond detectors can reach unexplored parameter space below stellar emission constraints, probing the QCD axion itself at masses ${\cal O}(10\;{\rm eV})$.

\subsection{Electron-Recoiling Dark Matter}\label{sec:erdm}
We now present the expected sensitivity of diamond detectors to electron recoils sourced by DM scattering. 
The event rate due to dark matter particle scattering off electrons is given by~\cite{Essig:2015cda}
\begin{equation}
\label{eq:rateER}
    \frac{dR}{d\ln E} = \frac{\rho_{\chi}}{m_{\chi}}\frac{m_{\rm det}}{m_{\rm cell}}\bar{\sigma}_e\alpha\frac{m_e^2}{\mu_{\chi e}^2} I_{\rm crystal}(E;F_{\rm DM})
\end{equation}
where $m_{\rm det}$ is the detector mass, $m_{\rm cell}$ is the mass of a single unit cell of the substrate, $\bar{\sigma}_e$ is the (reference) cross-section on electrons, $\alpha$ is the fine-structure constant, $\mu_{\chi e}$ is the electron-DM reduced mass, and $I_{\rm crystal}$ is the integrated form factor defined in Ref.~\cite{Essig:2015cda}.

\begin{figure*}
    \includegraphics[width=3.4in]{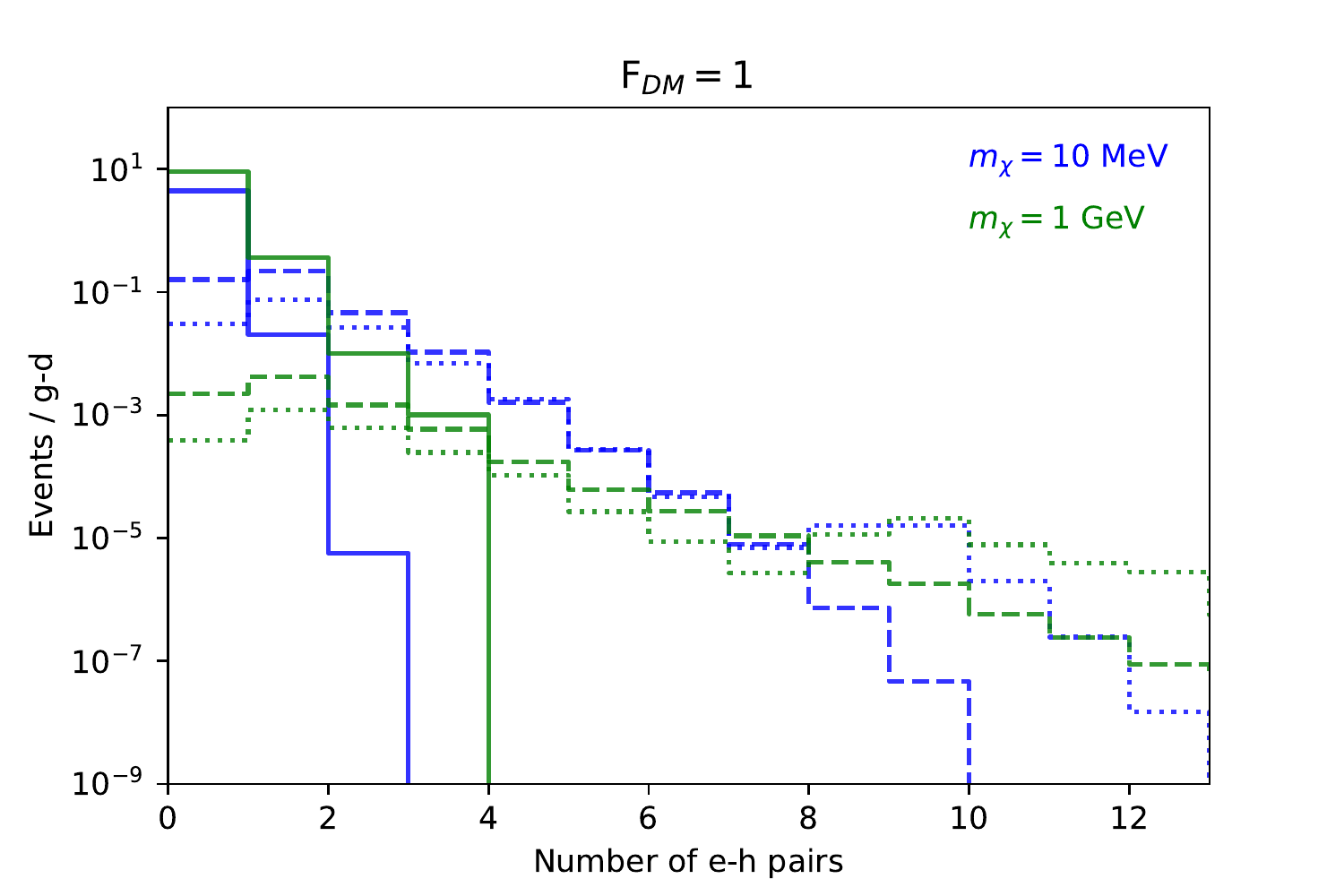}
    \includegraphics[width=3.4in]{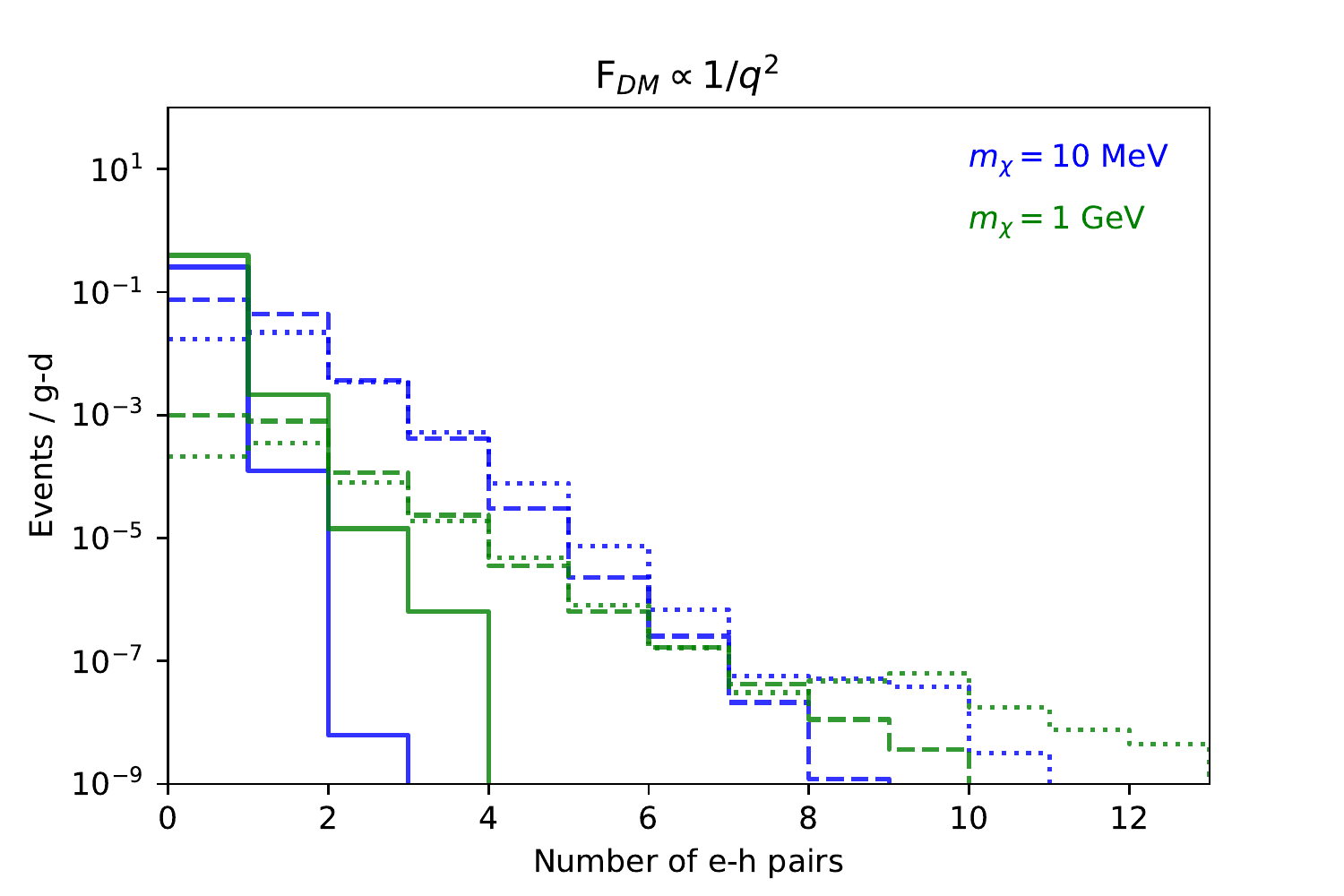}
    \caption{Recoil spectra for $\bar{\sigma}_e=10^{-37}$~cm$^2$ of silicon (dashed), germanium (dotted) and diamond (solid) for $m_{\chi}=$ 10 MeV (blue) and 1 GeV (green) respectively.}
    \label{fig:ER_spectra}
\end{figure*}

The integrated form factor is calculated numerically with the \texttt{QEDark}~\cite{Essig:2015cda} package. We ran the diamond calculation on a 8 $k$-point grid. The resulting recoil spectra are plotted in Fig.~\ref{fig:ER_spectra} for two different DM form factors. Due to the high band gap and high electron-hole pair energy in diamond, the energy threshold is higher and significantly fewer electron-hole pairs are produced for each event. As such, single e-h pair resolution becomes a necessary condition for a DM search using the electron recoil channel. Once this condition is met, however, the overall event rate for masses producing recoils above the bandgap energy will be higher than silicon and germanium due to the higher electron density in diamond. 

In fact, resolving single e-h pair may not be as technically challenging in diamond compared to silicon or germanium as the breakdown voltage in diamond is much higher than in the other materials. We can easily boost the energy of single e-h pair past our sensor threshold by applying a high voltage across the crystal. In addition, the absence of shallow impurities in diamond and the high band gap suggests a lower background rate. As such it should not be too difficult to achieve single e-h pair resolution in diamond without compromising signal-to-noise ratio.

\begin{figure*}
    \centering
    \includegraphics[width=3.4in]{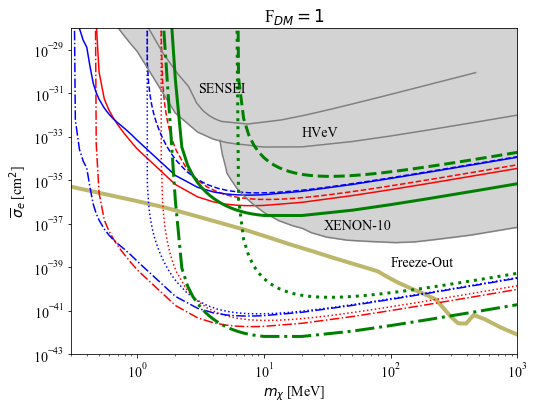}
    \includegraphics[width=3.4in]{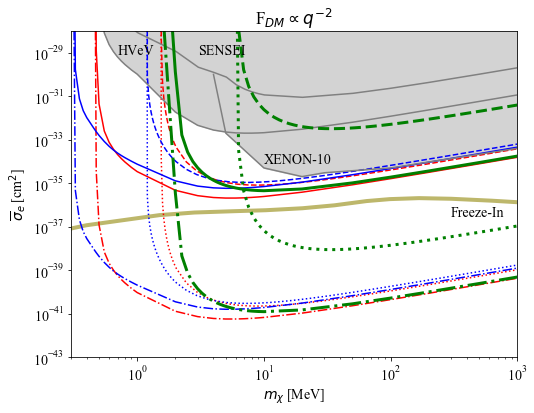}
    \caption{Projected reach at 95\% C.L. for electron recoils in silicon (red), germanium (blue) and diamond (green) detectors with 1 g-day of exposure and 1 e-h pair threshold (solid), 1 g-day of exposure and 2 e-h pair threshold (dashed), 1 kg-year of exposure and 1 e-h pair threshold (dot-dashed), and 1 kg-year of exposure and 2 e-h pair threshold (dotted). The current exclusion limits from SENSEI~\cite{sensei}, CDMS HVeV~\cite{Agnese:2018col}, and XENON10~\cite{Essig:2012yx} are also depicted for comparison. The parameter space corresponding to producing the observed dark matter relic density via standard freeze-out in a minimal dark sector model is indicated by the solid thick mustard curve labeled `Freeze-Out' (see Ref.~\cite{cosmicVisions} for more details).}
    \label{fig:ER_limit}
\end{figure*}

We show the 95\% C.L. projected reach, corresponding to 3 signal events, for a diamond electron-recoil detector in Fig.~\ref{fig:ER_limit} for both single e-h pair and 2 e-h pair thresholds. The current exclusion limits from SENSEI~\cite{sensei}, CDMS HVeV~\cite{Agnese:2018col}, and XENON10~\cite{Essig:2012yx} are also plotted for comparison.

\subsection{Nuclear-Recoiling Dark Matter}\label{sec:nr}
One of the distinct advantages of diamond is its high crystal purity, and the relatively few stable isotopes which exist in nature. Of these, $\mathrm{C^{12}}$ is the most naturally abundant, but crystals made of predominantly $\mathrm{C^{13}}$ are easily grown with enriched seed material, and produce crystals with nearly identical physical properties, with the exception of a nucleus capable of spin-dependent interactions~\cite{Neves2001}. This strongly implies that diamond is the preferred material for low-mass, spin-dependent DM searches. Here we only consider spin-independent limits on the DM-nucleon cross-section, deferring spin-dependent projections to future work, and are thus insensitive to the particular isotopic makeup of a given detector.

\begin{figure*}[t]
    \centering
    \includegraphics[width=3.4in]{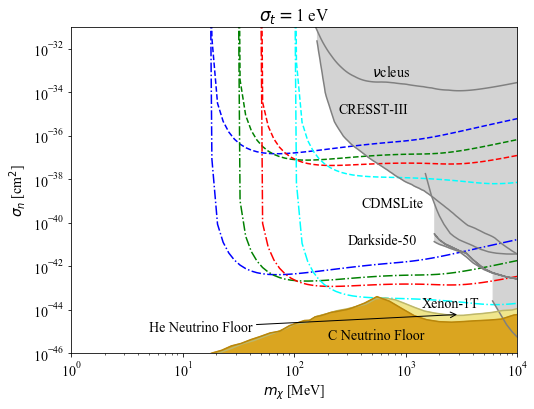}
    \includegraphics[width=3.4in]{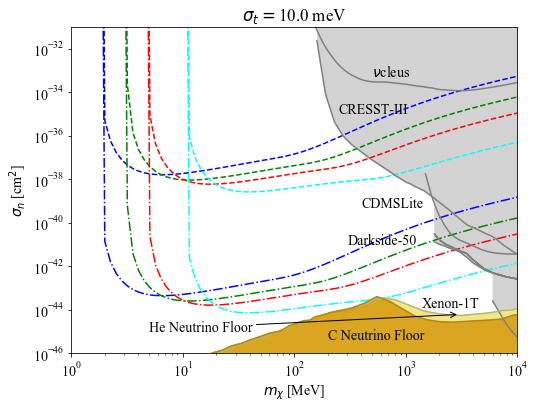}
    \caption{Nuclear recoil projected reach at 95\% C.L.  for He (blue), Diamond (green), Si (red), and Xe (cyan) for energy thresholds of 1~eV ({\it left panel}) and 10~meV ({\it right panel}). The dashed lines are for a g-day exposure, while the dot-dashed lines are for a kg-year exposure. The yellow region indicates the neutrino floor for He~\cite{hertel}, and the brown region the neutrino floor for C (computed according to the formalism in Ref.~\cite{billard}). Also shown in grey are the current best limits on NR dark matter interactions from $\nu$CLEUS~\cite{nucleus}, CRESST-III~\cite{CRESSTIII,CRESSTIIIIDM}, CDMSLite~\cite{cdmsliteR3}, Darkside-50~\cite{Darkside50}, and Xenon-1T~\cite{Xenon1T} for comparison.}
    \label{fig:NRlimits}
\end{figure*}

The event rate due to dark matter scattering off of a nucleus, in the spin-independent case, is given by the standard expression~\cite{lewinsmith}
\begin{equation}
\frac{dR}{dE_r} = m_{\rm det}\frac{\rho_{\chi}\sigma_0}{2m_{\chi}\mu_{\chi}^2}F^2(E_r)\int_{v_{\rm min}}^{v_{\rm esc}}\frac{f(v)}{v}d^3v,
\end{equation}
where $\mu_{\chi}$ is the reduced mass of the DM-nucleus system, $F(E_r)$ is the nuclear form factor of DM-nucleus scattering (we adopt the Helm form factor as in Refs.~\cite{lewinsmith,billard}), and $f(v)$ is the Maxwell-Boltzmann velocity distribution with parameters given in the beginning of this section. The cross-section $\sigma_0$ is normalized to a target nucelus, but to compare different media, this cross-section is re-parameterized as~\cite{lewinsmith,hertel}
\begin{equation}
\sigma_0 = A^2\left(\frac{\mu_{\chi}}{\mu_{\chi,n}}\right)^2\sigma_{n},
\end{equation}
where $A$ is the number of nucleons in the nucleus, and $\mu_{\chi,n}$ is the DM-nucleon reduced mass.  

For a sub-GeV dark matter particle, we find $\mu_{\chi}\rightarrow m_{\chi}$, $\sigma_0\rightarrow A^2\sigma_n$, and $F(E_r)\rightarrow 1$, such that
\begin{equation}
\frac{dR}{dE_r} \approx m_{\rm det}\frac{\rho_{\chi}A^2\sigma_n}{2m_{\chi}^3}\int_{v_{\rm min}}^{v_{\rm esc}}\frac{f(v)}{v}d^3v,
\end{equation}
which would seem to imply that a heavier nucleus is always more sensitive to dark matter from a pure event-rate perspective. Hidden in the integral, however, is the fact that
\begin{equation}
v_{\rm min} = \sqrt{\frac{E_r(m_{\chi}+m_T)}{2\mu_{\chi}m_{\chi}}} \rightarrow \sqrt{\frac{E_R m_{T}}{2m_{\chi}^2}}
\end{equation}
in this limit, which implies that there is a range of masses for which scattering off of heavier targets is kinematically suppressed. Thus detectors made of lighter nuclei will have better sensitivity to dark matter in this mass range, and be affected by this suppression at lighter masses. For this reason, hydrogen, helium, and carbon-based targets are being explored as sub-GeV dark matter detection media, in contrast to the dominance of heavy elements for the high-mass DM searches underway.

To compute NR limits, unlike in the electron-recoil case, the low-mass behavior is strongly dependent on the energy threshold, while the high-mass behavior depends on the upper limit for accurate energy reconstruction. TES-based calorimeters can provide very low thresholds but are intrinsically limited in dynamic range. To account for this, we assume 3 orders of magnitude in dynamic range, similar to what has been seen in detectors with O(eV) thresholds~\cite{kurinsky}. This means that the upper integration limit is set to $10^3\sigma_{t}$, where the threshold $\sigma_t$ is assumed to be 5 times the resolution. 

We show the 95\% C.L. projected reach, corresponding to 3 signal events, for calorimetric diamond detectors with the thresholds discussed in the previous section in Fig.~\ref{fig:NRlimits}, compared to the leading low-mass NR limits from the $\nu$cleus (sapphire, Ref.~\cite{nucleus}), CRESST-III ($\mathrm{CaWO_4}$, Ref.~\cite{CRESSTIII}), and CDMSlite (Ge, Ref.~\cite{cdmsliteR2}) experiments. Also shown is the neutrino floor calculated for He~\cite{hertel} and C (computed according to the formalism in Ref.~\cite{billard}). These projections demonstrate that even a 30~mg detector operated for a day at a surface facility covers previously unexplored parameter space, and operating for a month with a moderately low threshold can cover orders of magnitude of new parameter space. kg-year exposures bring the reach of diamond detectors near the neutrino floor, and would require significant background mitigation, and represent a large-scale experiment with costs and complexity on the order of currently operating GeV-scale dark matter searches.

\section{Discussion}\label{sec:disc}

In this work, we have demonstrated that diamond has significant reach for nuclear-recoil, electron-recoil and absorption detection channels, and is able to compete with traditional semiconductors in charge resolution, as well as with superfluid He in the nuclear recoil space due to its low atomic mass and long-lived phonon excitations. The synergy between diamond's potential as a dark matter detector and the potential application of diamond detectors to coherent neutrino scattering and UV imaging makes a compelling case for developing general purpose cryogenic diamond detectors.

A research program to demonstrate proof of principle for these designs is currently underway. We hope to report, in the near future, the successful fabrication of QET arrays on a high-purity diamond substrate, and to measure the phonon collection efficiency of these sensors on single and polycrystalline substrates. One significant advantage of diamond is its inertness relative to Si and Ge, which means that we should be able to apply the same fabrication techniques to diamond as our normal detector substrates, and the same set of tools can be used to quickly make these proof-of-principle devices. Beginning the development on polycrystalline diamond, and subsequently moving to larger crystals, will also help  determine the role that boundary scattering plays in phonon propagation and down-conversion. The path towards gram-year exposures, and research leading to lower thresholds and low-energy diamond tracking detectors, fits well within the scope of a small, early-phase experimental program of the type currently being explored to push to lower dark matter masses.

A significant barrier to scaling this technology to kg-year exposures is the cost of purchasing sufficient quantities of diamond substrates. While this has traditionally been the case, significant progress in CVD diamond growth driven by both the electronics industry and investment from quantum computing initiatives has made artificial diamonds now significantly less expensive than natural diamond of comparable quality. The remaining barriers to wide-scale adoption of diamond in research and technology are now mainly sociological, as natural diamond producers determine how to maintain a separation between natural and synthetic diamond in order to protect their investment (see {\it e.g.} Ref.~\cite{labdiamond}). 

There is a general consensus that this barrier will be overcome in the coming years, as has already been the case for sapphire, a similarly precious crystal which has seen wide-scale adoption in industry, and in its natural form is still extremely financially lucrative. DeBeers recently began selling lab-grown CVD diamond at a price point that would make 1~kg of diamond the same cost as the Xenon procured for LZ~\cite{diamondPrice} at current market price, while the infrastructure costs for diamond would be drastically less. If this trend continues, a kg-scale diamond experiment is likely to be well within the budget of a small-scale experiment. In addition, if it is determined that polycrystalline diamond has sufficient transport properties to achieve the resolutions described here, the cost of diamond substrates would be substantially lower than single crystal substrates, allowing for 10s of kg in fiducial mass at the same cost.

One interesting comparison noticed during the compilation of the dark matter absorption limits is the relative strength of the diamond, GaAs and sapphire absorption limits around the $\sim$100~meV energy range. As noted in Ref.~\cite{Knapen:2017ekk}, polar materials allow for enhanced absorption of sub-gap photons due to the high polarizability of their substrates, which allows single phonons to be produced from single photons in an inelastic interaction. Diamond, on the other hand, is non-polar, meaning that absorption can still happen, but requires the pair-production of phonons, reducing the cross-section by an extra coupling factor. 

In this context, we note another material that has seen recent work, SiC, which has very similar charge and phonon properties to diamond, but is also polar; the implications are that it would be able to take advantage of the polar phonon scattering mechanism (see {\it e.g.} Ref.~\cite{siliconCarbide}). Compared to diamond, its smaller bandgap of order $\sim$3.2~eV along with its high-energy optical phonon modes, make a large, pure sample likely to have slightly better reach in the electron-recoil space, and slightly less reach in the nuclear recoil space. In all other respects, it should behave as an intermediate material between Si and diamond.  The full exploration of SiC as a target material is beyond the scope of this work; we do not consider it further here due to the high likelihood of more complex impurity structures, in addition to the likelihood that phonon scattering lengths will be much shorter due to the mixed atomic composition.
The prospects of SiC and other carbon-based compound semiconductors will be explored in detail in a future paper.

\begin{acknowledgements}
We would like to thank Dan Bauer, Paul Brink, Scott Hertel, Lauren Hsu, Matt Pyle, and Kathryn Zurek for useful discussions about the feasibility of diamond as a detector material, Rouven Essig and Tien-Tien Yu for help with \texttt{QEDark} and advice related to setting electron-recoil limits, Jason Pioquinto and Tarek Saab for neutrino floor calculations in Carbon, and Paolo Privitera and Yoni Kahn for discussions related to $^{14}$C backgrounds. We also thank Alissa Monte for detailed comments on an early draft of this paper, and Betty Young for subsequent feedback.  The work of YH is supported by the Israel Science Foundation (grant No. 1112/17), by the Binational Science Foundation (grant No. 2016155), by the I-CORE Program of the Planning Budgeting Committee (grant No. 1937/12), by the German Israel Foundation (grant No. I-2487-303.7/2017), and  by the Azrieli Foundation. TCY acknowledges support by the U.S. Department of Energy, Office of Science, Office of HEP. This document was prepared by NK using the resources of the Fermi National Accelerator Laboratory (Fermilab), a U.S. Department of Energy, Office of Science, HEP User Facility. Fermilab is managed by Fermi Research Alliance, LLC (FRA), acting under Contract No. DE-AC02-07CH11359.
\end{acknowledgements}

\bibliographystyle{revtex-4-1.bst}
\bibliography{refs.bib}

\end{document}